\documentclass[a4paper,11pt]{article}
\usepackage{pos}
\usepackage{braket}
\usepackage{subcaption}

\RequirePackage{color}
\definecolor{dkgreen}{rgb}{0.2,0.7,0.4}
\definecolor{dkblue}{rgb}{0.2,0.2,0.7}
\definecolor{dkred}{rgb}{0.8,0,0}
\definecolor{dkpurple}{rgb}{0.45,0.2,0.55}

\title{Spectrum-Generating Algebra in Higher Dimensional Gauge Theories}

\author[a]{Thea Budde}
\author[a,b]{Jiangjing Dong}
\author[a]{Marina K. Marinkovi\'{c}}
\author*[a]{Joao C. Pinto Barros}

\affiliation[a]{Institut f\"{u}r Theoretische Physik, ETH Z\"{u}rich,\\
  Wolfgang-Pauli-Str. 27, 8093 Z\"{u}rich, Switzerland}

\affiliation[b]{Institute for Theoretical Physics, KU Leuven,\\
Celestijnenlaan 200D, 3001 Leuven, Belgium}

\emailAdd{jpinto@phys.ethz.ch}

\abstract{Non-equilibrium properties of strongly interacting gauge theories are often intractable with classical simulation methods. Due to recent developments of quantum simulations, studies of their properties in two spatial dimensions are becoming accessible. By demonstrating the existence of an approximate spectrum-generating algebra for a pure gauge plaquette ladder, we predict and verify the existence of Quantum Many-Body Scars in spin-1 Quantum Link Models. The analysis of the model is facilitated by a dualization process that maps the original gauge theory to a constrained spin chain. Was it not for the constraint, the system would have an exact spectrum-generating algebra. We propose a set of observables for diagnosing an approximate spectrum-generating algebra, which is expected to guide quantum simulators toward interesting physical regimes.}

\FullConference{
}

\begin{document}
\maketitle

\section{Introduction and the Model}

The prospect of realizing gauge theories in quantum simulators has sparked significant interest in their non-equilibrium properties. For some time, short- and medium-time dynamics have held the promise of being the most interesting regime to study and characterize, while long-time dynamics were expected to thermalize, according to the Eigenstate Thermalization Hypothesis (ETH) \cite{srednickiApproachThermalEquilibrium1999,dalessioQuantumChaosEigenstate2016}. The discovery of Quantum Many-Body Scars (QMBS) has challenged this picture \cite{bernienProbingManybodyDynamics2017}. In particular, gauge theories seem very prone to admit QMBS in their spectrum and evade ETH in general \cite{banerjeeQuantumScarsZero2021,biswasScarsProtectedZero2022,desaulesProminentQuantumManybody2023,halimehRobustQuantumManybody2023,desaulesWeakErgodicityBreaking2023,geNonmesonicQuantumManyBody2024,sauSublatticeScarsTwodimensional2024,buddeQuantumManybodyScars2024,calajoQuantumManybodyScarring2025,hartseStabilizerScars2025,miaoExactQuantumManyBody2025,guptaExactStabilizerScars2026,Budde2026}. Although different mechanisms for these phenomena have been described (e.g., \cite{moudgalyaQuantumManybodyScars2022,chandranQuantumManyBodyScars2023,pakrouskiGroupTheoreticApproach2021a}), it remains a challenge to determine whether a given generic model exhibits abnormal thermalization properties. Another interesting question is the interplay between gauge symmetry and the emergence of quantum many-body scars. Although it is clear that one does not imply the other, it would be interesting to understand why QMBS seem to be so abundant in these models. Here, we focus on a pure-gauge theory in quasi-one dimension. We will focus on the spin-1 Quantum Link Model (QLM), which consists of a plaquette ladder. This lies at the edge of what can be achieved with near-term quantum simulators \cite{daiFourbodyRingexchangeInteractions2017,fontanaQuantumSimulatorLink2023,cochranVisualizingDynamicsCharges2025,methSimulatingTwodimensionalLattice2025,muellerQuantumComputingUniversal2025}. The model also exhibits a rich spectrum, which leads to anomalous thermalization due to low-entanglement states in the middle of the spectrum, for arbitrary (even) volumes \cite{buddeQuantumManybodyScars2024}. We build on \cite{buddeQuantumManybodyScars2024} and show that the spectrum also hosts an approximate spectrum-generating algebra, leading to persistent revivals from low-entropy initial states.

\begin{figure}
    \centering
    \begin{subfigure}{.29\linewidth}
        \centering
        \includegraphics[width=\linewidth]{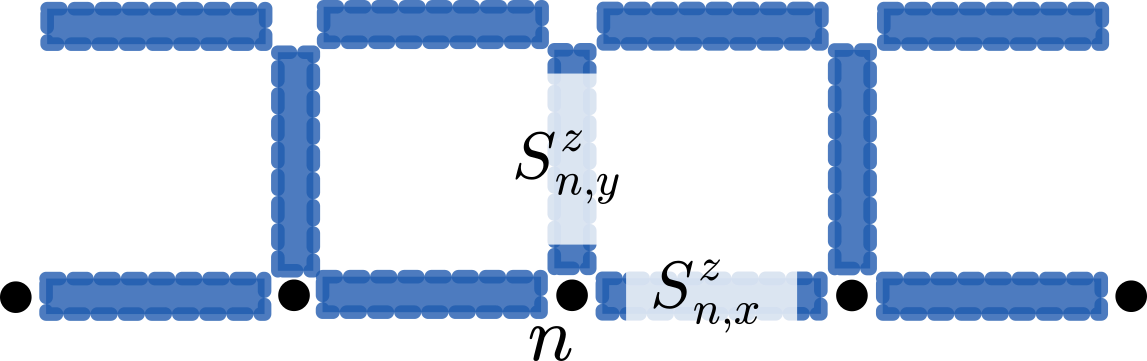}
        \caption{Original formulation with spin-1 variables $S^a_{n,i}$ defined on the links $(n,i)$ of a ladder.}
        \label{fig:ladder_QLM}
    \end{subfigure}
    \hspace{.03\linewidth}
    \begin{subfigure}{0.29\linewidth}
        \centering
        \includegraphics[width=\linewidth]{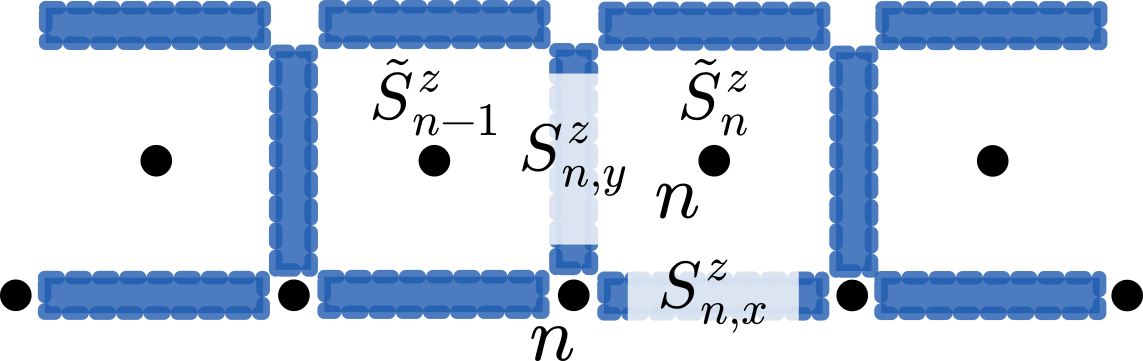}
        \caption{Original variables $S^a_{n,i}$ determine the dual spin variables $\tilde{S}^a_{n}$ residing at the center of plaquettes.}
        \label{fig:ladder_dualization}
    \end{subfigure}
    \hspace{.03\linewidth}
    \begin{subfigure}{0.29\linewidth}
        \centering
        \includegraphics[width=\linewidth]{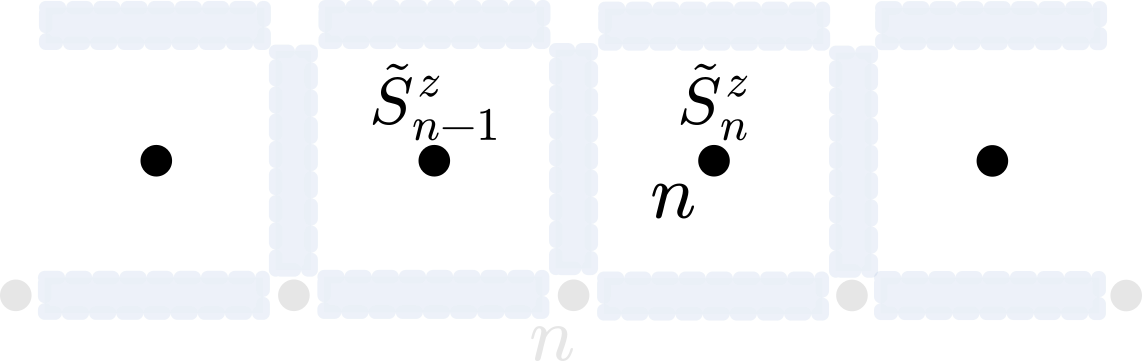}
        \caption{The dualized model is effectively described by a spin-1 chain, constrained according to Eq. \eqref{eq:H_constrained}.}
        \label{fig:ladder_dual}
    \end{subfigure}
    \caption{Process of dualization: from the original gauge theory to the constrained spin chain.}
    \label{fig:dualization_process}
\end{figure}

We make use of the quasi-one-dimensional nature of the model and interchangeably denote by $n$ the lowest vertices of the ladder identified by a single integer or by a two-component vector, i.e., $n\equiv\left(n,1\right)$. We denote the unit vectors $\hat{x}=(1,0)$ and $\hat{y}=(0,1)$. With this notation, we will be interested in Hamiltonians of the form

\begin{equation}
    H = \sum_{n=1}^L \left( S^+_{n,x} S^+_{n+\hat{x},y} S^{-}_{n+\hat{y},x} S^-_{n,y} + \text{h.c.}\right) + V.
    \label{eq:Hpure_gauge}
\end{equation}
The spin-raising and spin-lowering operators correspond to spin-1 variables. The sum corresponds to the so-called ``plaquette terms'' in a ladder of of $L$ plaquttes. An illustration of the geometry of the model can be found in Fig. \ref{fig:ladder_QLM}. Our results will focus on $V=0$ and we use periodic boundary conditions. The relevant symmetries for the present discussion are the following.
\begin{enumerate}
    \item \emph{gauge symmetry}, where $\left[H,G_n\right]=0$ and $\left[H,G_{n+\hat{y}}\right]=0$ for all $n$, with $G_n=S^z_{n,x}-S^z_{n-\hat{x},x}+S^z_{n+\hat{y},y}$ and $G_{n+\hat{y}}=S^z_{n+\hat{y},x}-S^z_{n+\hat{y}-\hat{x},x}-S^z_{n,y}$,
    \item \emph{winding symmetries}, which, in the ladder geometry chosen here, correspond to the conservation of two types of magnetization $w_y=S^z_{n,x}+S^z_{n+\hat{y},x}$ and  $w_x=\sum_nS^z_{n,y}$,
    \item \emph{translation invariance} in the $\hat{x}$ direction characterized by the transformations $S^z_n\rightarrow S^z_{n+m\hat{x}}$ with $m\in\mathbb{Z}$,
    \item \emph{reflection symmetry} with respect to the $y$ axis, where $S^z_{n,1}\rightarrow-S^z_{(L+1)\hat{x}-n,x}$ and $S^z_{n,2}\rightarrow S^z_{(L+1)\hat{x}-n,y}$.
\end{enumerate}

We consider only the physical Gauss' law and zero-winding sectors defined by states $\ket{\psi}$ that satisfy the conditions
\begin{equation}
    G_n\ket{\psi}=0,\quad G_{n+\hat{y}}\ket{\psi}=0,\quad w_x\ket{\psi}=0,\quad w_y\ket{\psi}=0\quad\quad\forall n.
    \label{eq:hilbert_space}
\end{equation}
The last equation immediately implies $S^z_{n,x}\ket{\psi}=-S^z_{n+\hat{y},x}\ket{\psi}$, so we only need one horizontal spin per-row. Furthermore, if we know two consecutive horizontal links in the basis where $S^z$ are diagonal, we can uniquely determine the value of the spin on the vertical link between them. This results in a mapping of a ladder geometry to a spin chain of spin-1 variables, denoted by $\tilde{S}^a_n$ with $n\in\left\{1,\dots, L\right\}$. This is a specific instance of the dualization process, illustrated in Fig. \ref{fig:dualization_process}. The more general construction can be found in \cite{buddeQuantumManybodyScars2024}. As it turns out, we can identify the dual variables with the value of the upper horizontal spin as $\tilde{S}^z_n\rightarrow S^z_{n+\hat{y},x}$ and the value of the vertical spins recovered has the difference of neighboring height variables $S^z_{n,y}=\tilde{S}^z_n-\tilde{S}^z_{n-1}$.
After the described mapping to a spin chain we can write the effective Hamiltonian in Eq. \eqref{eq:Hpure_gauge} in terms of dual variables
\begin{equation}
    H=\sum_n{\cal P}_n\tilde{S}^x_n{\cal P}_n+\tilde{V},
    \label{eq:H_constrained}
\end{equation}
where $\tilde{V}$ is the corresponding dualized potential $V$ and the plaquette terms are encoded in the first sum, making use of the projectors ${\cal P}_n$, which forbid neighboring dual spins to be in states $-1$ and $+1$, of the $z$-basis (i.e. where $\tilde{S}^z_n$ are diagonal). Explicitly, we can write ${\cal P}_n = P^{(0)}_{n} +Q^{(1)}_{n-1}P^{(-1)}_{n}Q^{(1)}_{n+1} + Q^{(-1)}_{n-1}P^{(1)}_{n}Q^{(-1)}_{n+1}$, where $Q^{(s)}=1-P^{(s)}$ and $P^{(s)}$ is the projector in the spin state $s$. On the $z-$basis, the projectors act according to $P^{(s)}\ket{s^\prime}=\delta_{ss^\prime}\ket{s^\prime}$. Note that ${\cal P}_n$ has support on three sites: $n-1$, $n$ and $n+1$. This dualization process is valid when we restrict ourselves to the constrained Hilbert space where ${\cal P}_n\ket{\psi}=\ket{\psi},\ \forall n$.
Within the target Hilbert space defined in Eq. \eqref{eq:hilbert_space}, the two Hamiltonians in Eqs. \eqref{eq:Hpure_gauge} and \eqref{eq:H_constrained} are equivalent. The latter form will be particularly useful for studying the existence of QMBS from the perspective of an approximate spectrum-generating algebra, which we discuss in the next section, similar to what has been found for the PXP model \cite{bullQuantumScarsEmbeddings2020}.

The rest of this proceedings is organized as follows. In Sec. \ref{sec:sga}, we review the concepts of spectrum-generating algebras and broken Lie algebras. The latter occurs when the system does not possess an exact dynamical symmetry but an approximate one. In Sec. \ref{sec:spectral_analysis}, we analyze the spectrum of the Hamiltonian in terms of its entanglement entropy and a newly introduced observable, the broken Casimir. In Sec. \ref{sec:tevolution}, we make informed guesses of initial states that can show a lack of thermalization and we conclude in Sec. \ref{sec:conclusion}.

\section{Spectrum-Generating Algebra and Broken Lie Algebras\label{sec:sga}}

In free spin systems, one finds a spectrum-generating algebra (or dynamical symmetry)\cite{moudgalyaQuantumManybodyScars2022}, which amounts to the existence of an operator $O^\dagger$ that satisfies the commutation relation with the Hamiltonian $\left[H,O^\dagger\right]=\omega O^\dagger$. This means that $O^\dagger$ acts as a raising operator for the eigenstates, so that if $\ket{E}$ is an eigenstate with energy $E$, then $O^\dagger \ket{E}$ is either a null state or an eigenstate with energy $E+\omega$. Such an algebraic structure allows the construction of towers of states, spaced by energy $\omega$. Any initial state constructed from states belonging to a single such tower will exhibit perfect revivals under time evolution \cite{moudgalyaQuantumManybodyScars2022}. The existence of these towers is more surprising when considering interacting models and entails several examples that include the Hubbard model \cite{yangPairingOffdiagonalLongrange1989,yangSO4SYMMETRYHUBBARD1990}, the Spin-1 $XY$ model \cite{schecterManybodySpectralReflection2018} and the AKLT model \cite{moudgalyaExactExcitedStates2018}. 

Interestingly, it is also possible to observe approximate revivals in models that do not possess an exact dynamical symmetry, but rather an approximate one, or a \emph{broken Lie Algebra}. In this framework, an operator $O^\dagger$ that satisfies $\left[H,O^\dagger\right]=\omega O^\dagger$ does not exist, but there exist some $O^\dagger$ where the relation is approximately valid for a family of eigenstates. They do not form a Lie algebra but instead a ``broken Lie algebra''. For this series of eigenstates, we sometimes call \emph{broken towers} as they exhibit an approximate constant energy spacing. This was found to be the case for the PXP model \cite{bullQuantumScarsEmbeddings2020}. To make this discussion concrete, we address directly the free spin system derived from Eq. \eqref{eq:H_constrained} when the constraint is removed and consider, for simplicity, the case of $\tilde{V}=0$. In this case $H_0=\sum_nS^x_n$ and, together with $O^\dagger=\sum_n\left(S^y_n+i S^z_n\right)$ and $\omega=1$, they realize a dynamical symmetry. Furthermore, we have $O=\sum_n\left(S^y_n-i S^z_n\right)$ and the three operators $H$, $O$, and $O^\dagger$ furnish a representation of the $\mathfrak{su}(2)$ Lie algebra.

With the constraints included, this construction no longer works. In the next section, we will explore how part of the dynamical symmetry structure still survives in the spectrum of the Hamiltonian, leading to approximate towers and approximate revivals.

\section{Spectral Analysis: Entanglement Entropy and Broken Casimir \label{sec:spectral_analysis}}

\begin{figure}
    \centering
    \begin{subfigure}{.45\linewidth}
        \centering
        \includegraphics[width=\linewidth]{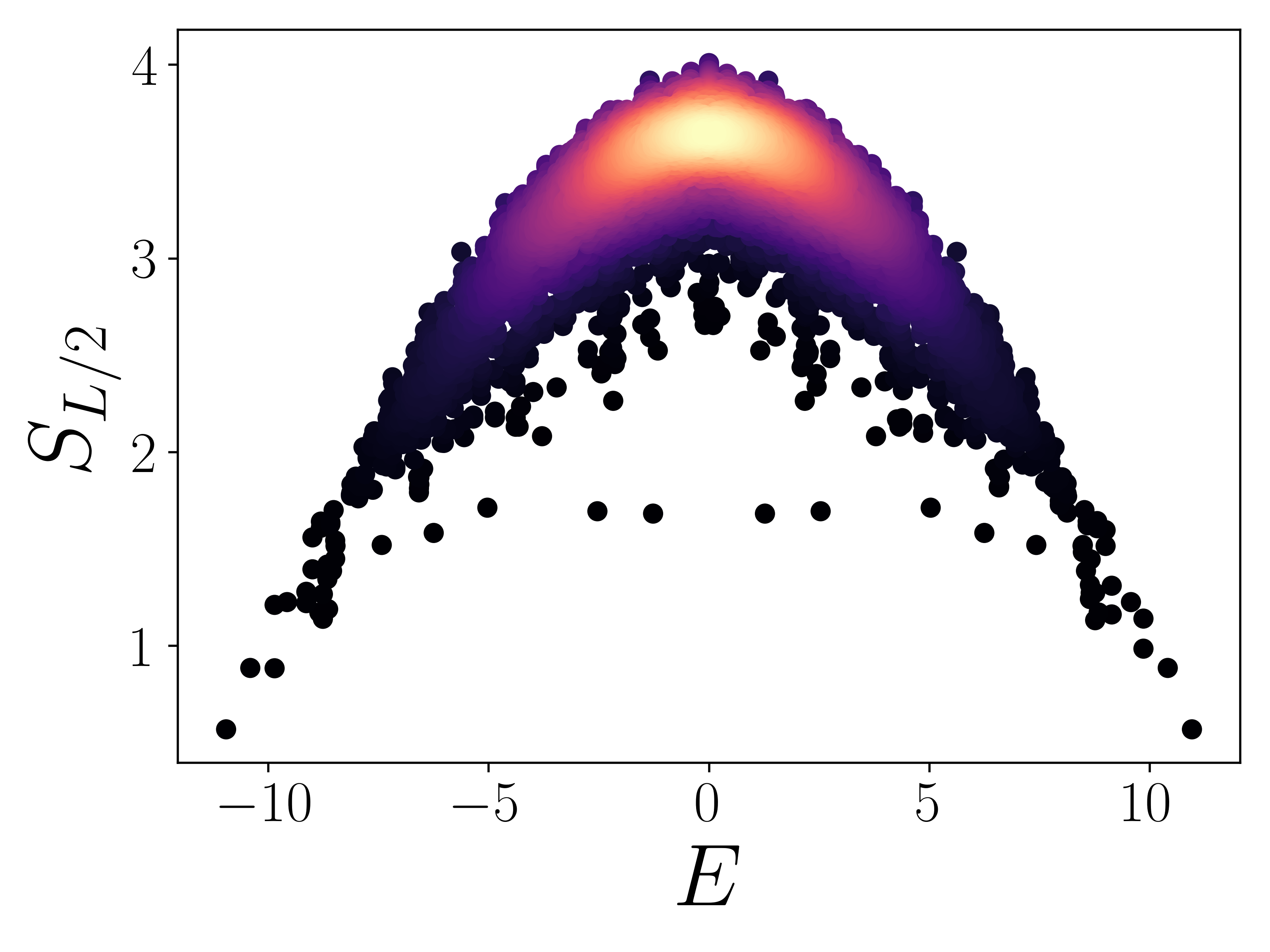}
        \caption{Entanglement entropy}
        \label{fig:entanglement}
    \end{subfigure}
    \hfill
    \begin{subfigure}{0.45\linewidth}
        \centering
        \includegraphics[width=\linewidth]{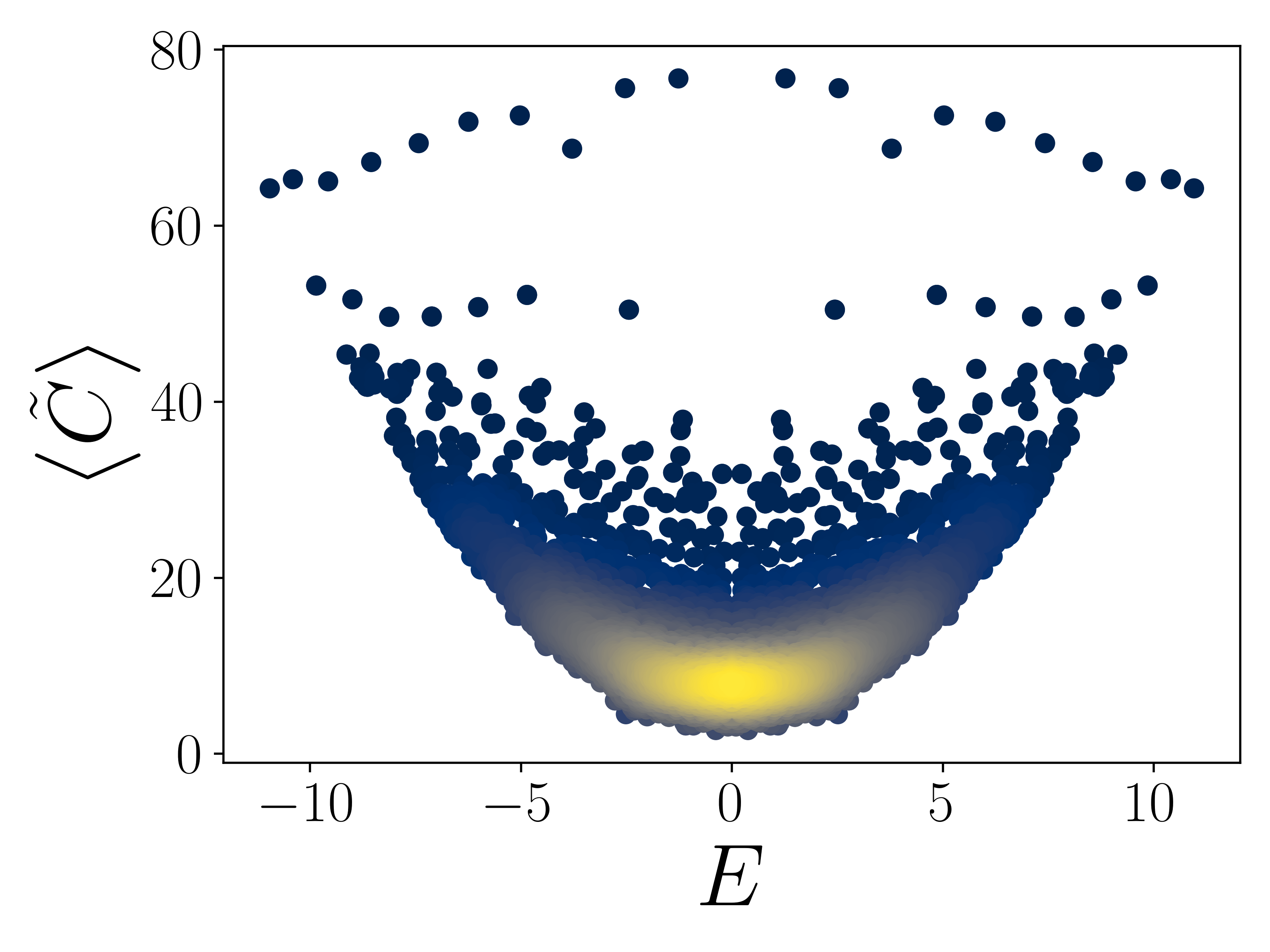}
        \caption{Broken Casimir}
        \label{fig:casimir}
    \end{subfigure}
    \caption{Entanglement entropy and expectation value of the broken Casimir operator in Eq. \eqref{eq:broken_casimir} for all eigenstates of the Hamiltonian. In both plots, we observe a family of outliers, seemingly with constant gaps between them, suggesting the existence of an approximate spectrum-generating algebra. We can further verify that the outliers are exactly the same states in both plots.}
    \label{fig:ab}
\end{figure}

In order to investigate whether the constraint allows for an approximate spectrum-generating algebra derived from a broken Lie algebra construction, we start by computing the half-system entanglement entropy. This is a useful diagnostic for the potential existence of QMBS, as mid-spectrum eigenstates of the Hamiltonian are expected to follow a volume-law for the entanglement entropy. States that appear as outliers at mid-spectrum in an entanglement entropy plot strongly hint at the possibility of abnormal thermalization from specific, simple, but fine-tuned initial states. For details of the computation of the half-system entanglement entropy, here denoted by $S_{L/2}$, we refer to the Supplemental Material of \cite{buddeQuantumManybodyScars2024}. In Fig. \ref{fig:entanglement} we plot the entanglement entropy for all the eigenstates as a function of the energy for $L=10$. We can identify a family of outliers that are seemingly equally spaced in energy, along with other states that lie outside the bulk of the eigenstates. Evidence of this structure could already be identified in \cite{buddeQuantumManybodyScars2024} with $\tilde{V}\neq0$, where other low-entanglement outliers were constructed analytically. 

In order to further probe the hypothesis that the outliers originate from a broken Lie algebra, we construct explicitly a candidate for such a broken Lie algebra. There are many ways to do this, but one minimal requirement is that it should become exact in the absence of the constraint. We define $\tilde{H}^+\equiv\sum_n{\cal P}_n\tilde{S}^+_n{\cal P}_n$ and propose the following set of operators
\begin{equation}
    \label{eq:broken_operators}
    \tilde{H}_x=\frac{1}{2}\left(\tilde{H}^++\tilde{H}^-\right),\ 
    \tilde{H}_y=\frac{1}{2i}\left(\tilde{H}^+-\tilde{H}^-\right),\ 
    \tilde{H}_z=\frac{1}{2}\left[\tilde{H}^+,\tilde{H}^-\right]. 
\end{equation}
This definition guaranties that $\tilde{H}_x$, is exactly our Hamiltonian in Eq. \eqref{eq:H_constrained} and that without the constraint, i.e., removing the projectors ${\cal P}_n$, this would form an exact $\mathfrak{su}(2)$ algebra where $\left[\tilde{H}^a,\tilde{H}^b\right]=\sum_c\varepsilon^{abc}\tilde{H}^c$. With the introduction of the constraints, these relations are no longer satisfied. It still might be possible that these relations can be approximately satisfied when restricted to a subspace of the full Hilbert space. We will see that this will be the case.

In order to test this hypothesis, we further define an observable that we call the \emph{broken Casimir}
\begin{equation}
\tilde{C}=\left(\tilde{H}_x\right)^2+\left(\tilde{H}_y\right)^2+\left(\tilde{H}_z\right)^2.
\label{eq:broken_casimir}
\end{equation}
If the constraining projectors were absent, this would be exactly the Casimir invariant, which would correspond to the total spin. The spectrum of eigenstates would be broken down into sectors with different total spin $j$ and eigenvalue of the Casimir operator $j(j+1)$. Each $j$ labels different towers, although several towers can share the same $j$. Using standard representation theory, we can count the number of different towers that exist for each $j$. We can now compute the expectation value of $\tilde{C}$ and check if any of these many towers survive approximately. This is depicted in Fig. \ref{fig:casimir}. The hint for an approximate tower should exhibit an approximately constant value of $\left<\tilde{C}\right>$, with states with approximately equal energy gaps. There are at least two sets of outliers that can be seen in Fig. \ref{fig:casimir} that follow this pattern, which correspond to the outliers in the entanglement entropy plot of Fig. \ref{fig:entanglement}. We should remark that in both sets, there seems to be a missing state for $E=0$. This happens because there is a very large degeneracy of eigenstates in $E=0$ \cite{schecterManybodySpectralReflection2018,buddeQuantumManybodyScars2024}. Under exact diagonalization, these states all mix arbitrarily and that state becomes invisible in the plot. \footnote{The state corresponding to $E=0$ can be made visible if we diagonalize the alternative Hamiltonian $H\rightarrow H+\varepsilon \tilde{C}$, making $\varepsilon$ very small, guaranteeing that the $E=0$ degeneracy is lifted, while most of the spectrum remains untouched.}

In addition to spotting two apparent towers, we can guess from which total spin $j$ they derive by simple state counting. For such towers, we expect $2j+1$ spins. From the states with higher values of $\left<\tilde{C}\right>$, we can count exactly 21 states (taking into account the hidden state at $E=0$) signaling $j=10$. For the second set, we count 15 (counting with the hidden $E=0$). This would suggest that this is a $j=7$ tower. However, there appear to be energy gaps that do not conform to the tower structure, near $E=\pm 2$ and $E=\pm 4$. If we assume the existence of hidden states in those places, we get $j=9$. The fact that we do not clearly see the missing states in $E=\pm 2$ and $E=\pm 4$ could be due to degeneracies or because the Lie algebra structure is more broken at those energies. We will see that $j=9$ is consistent with further analysis in the time dynamics.

\begin{figure}
    \centering
    \includegraphics[width=\linewidth]{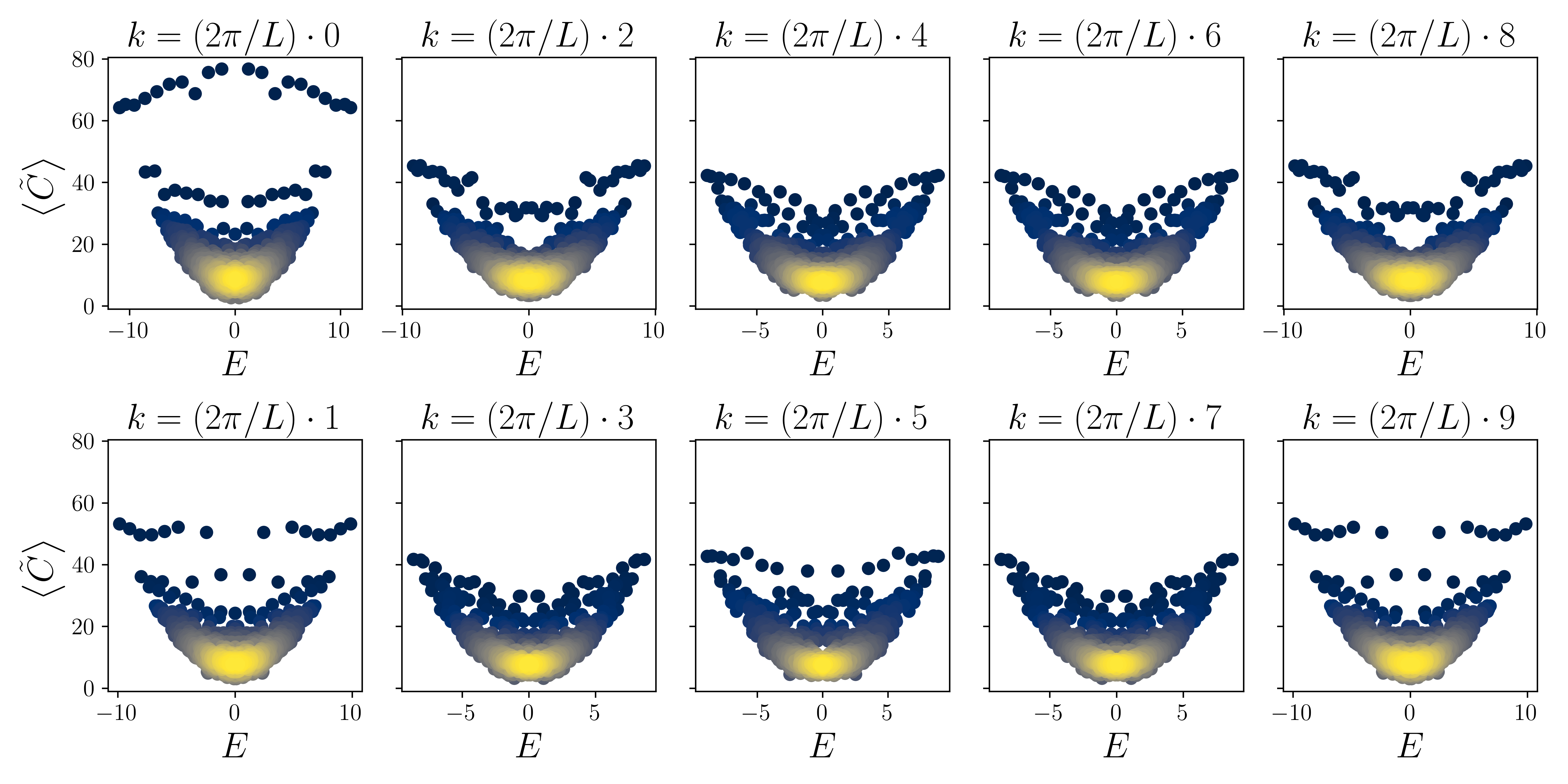}
    \caption{Expectation value of the broken Casimir for different eigenstates resolved in different momentum sectors. We observe that the most prominent tower in Fig. \ref{fig:casimir} lies in the zero-momentum sector, while the second one exists for $k=\pi/5$ and $k=9\pi/5$. Other approximate towers become visible. Most prominently at $k=\pi$ and a second at zero-momentum ($k=0$).}
    \label{fig:momentum_casimirs}
\end{figure}

Although there is only one tower, in the free-spin system, with $j=10$ (for volume $L=10$), there are $9$ different towers with $j=9$. We can identify which ones are surviving if we resolve the spectrum in momentum space\footnote{ Only possible due to translation invariance of the system.}. Momentum sectors can be labeled by the quantum number $k=\left(2\pi/L\right)m$ with $m\in\left\{0,\dots,9\right\}$. We plot the broken Casimir in all these sectors in Fig. \ref{fig:momentum_casimirs}.
There are several features that we can identify from these plots. First, the largest broken Casimir tower was completely moved to the $k=0$ sector, as expected. The second broken tower is visible in both $k=\left(2\pi/L\right)$ and $k=\left(2\pi/L\right)9$. These are, in fact, degenerate sectors that come from reflection symmetry, which maps the momentum sector $k$ to the $2\pi-k$ sector. The sector $k=\pi$ also seems to exhibit another approximate tower. Finally, $k=0$ seems to exhibit yet another approximate tower, below the most prominent one; however, from state counting, it should correspond to a $j=8$ irrep. and not $j=9$. We will not discuss this lower-spin tower here.

The broken Casimir allowed us to identify, numerically, a family of eigenstates that produce approximate towers of a broken Lie algebra. In the next Section, we will translate this into an analytical prediction of which initial simple states can produce revivals and verify it by computing the time evolution of the magnetization and the broken Casimir, initialized from those states.

\section{Scarred Initial States and Time-Evolution\label{sec:tevolution}}

If the Lie algebra were exact, we could observe exact revivals by initializing time-evolution from a state that can be written as a linear combination of states on a single tower. The towers can be identified by finding states $\ket{\psi_0}$, satisfying the equation $\tilde{H}^-\ket{\psi_0}=0$, and then repeatedly acting with $\tilde{H}^+$ to obtain the rest of the states in the tower. In other words, after identifying $\ket{\psi_0}$, the tower corresponds to states in the set $\left\{\left(\tilde{H}^+\right)^m\ket{\psi_0}\right\}_{m=0}^{2j}$, where $j$ is the total spin of the tower.

\begin{figure}
    \centering
    \includegraphics[width=\linewidth]{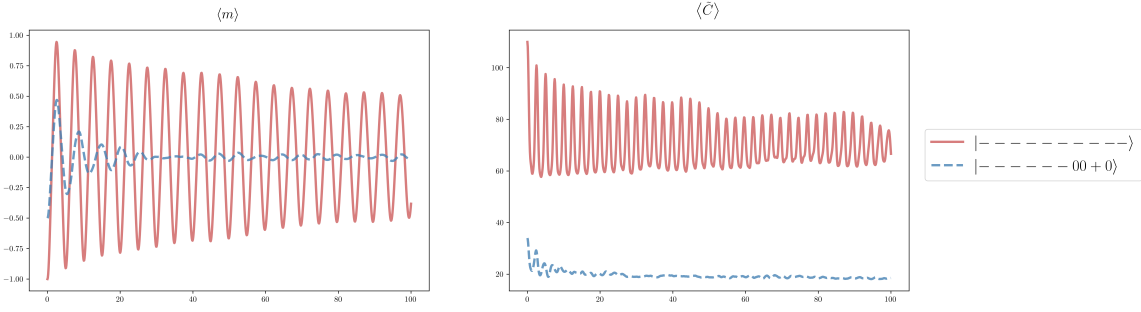}
    \caption{Time-evolution of the magnetization and the broken Casimir starting from the predicted scar state, $\ket{-}^{\otimes L}$ and a generic state $\ket{-}^{\otimes 6}\ket{0+00}$. The former, in contrast to the latter, exhibits persistent revivals of magnetization and abnormally large values of the broken Casimir for long times.}
    \label{fig:highest_tower}
\end{figure}

In Sec. \ref{sec:spectral_analysis}, we identified two potential towers with $j=10$ and $j=9$. For the former, there is only one candidate $\ket{\psi_0}$, corresponding to a product state with all spins in the state $-1$, in the $z-$basis, i.e. $S^z_n\ket{\psi_0}=-\ket{\psi_0},\ \forall n\in\{1,\dots,L\}$. We denote this state by 
\begin{equation}
    \ket{\phi_-}\equiv\ket{-}^{\otimes L}=\ket{----------}.
    \label{eq:phi_minus}
\end{equation}
To verify the existence of revivals, we track the time evolution of the broken Casimir and the magnetization, defined by $m=\left(1/L\right)\sum_n\tilde{S}^z_n$. We illustrate the time evolution starting from a state $\ket{\phi_-}$ and for an arbitrarily chosen generic state $\ket{-}^{\otimes 6}\ket{0+00}$ in Fig. \ref{fig:highest_tower}.
The predicted scarred state, $\ket{\phi_-}$, shows persistent (decaying) revivals in the magnetization, which can be traced back to the approximate broken Lie algebra structure uncovered in Sec \ref{sec:spectral_analysis}. The broken Casimir remains at an abnormally large value, especially when we note what is the expectation value for the bulk of most eigenstates, as depicted in Fig. \ref{fig:casimir}. Due to the non-locality of $\tilde{C}$, this does not qualify, per se, as a violation of ETH. However, it shows that the initial state exhibits an abnormal superposition with the outliers, characterized by a large broken Casimir.

\begin{figure}
    \centering
    \includegraphics[width=\linewidth]{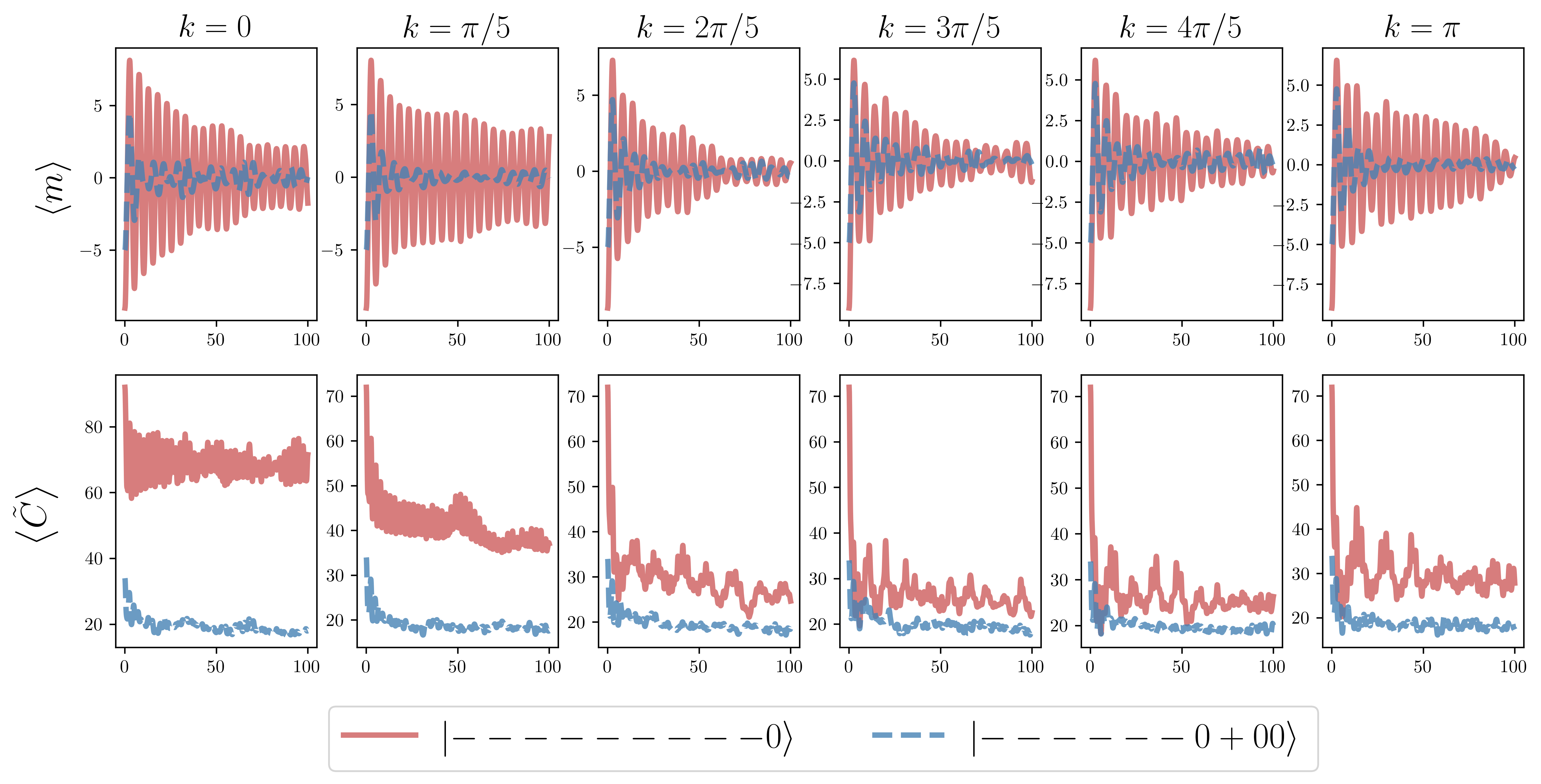}
    \caption{Time-evolution of the magnetization and broken Casimir for the projection of two product states for all momenta. As predicted the projection of $\ket{-}^{\otimes 9}\ket{0}$ (corresponding to $\ket{\phi_k}$ at $k=0$, $k=\pi/5$ and $k=\pi$ shows the most prominent revivals. Except for the latter, the broken Casimir retains abnormally large values throughout the time evolution.}
    \label{fig:all_momenta}
\end{figure}

We now turn our attention to the other towers. For $j=9$, we get $9$ towers, one for each momentum sector that is nonzero. These states can be constructed by placing a spin in the zero state, in a sea of $-1$'s and boosting it with momentum $k$. Explicitly, we define
\begin{equation}
    \ket{\phi_k}=\frac{1}{\sqrt{L}}\sum_{n=1}^Le^{ikn}S^+_n\ket{\phi_-}.
    \label{eq:phi_k}
\end{equation}
For $k=0$, there is no new tower. Rather we simply have $\sqrt{L}\ket{\phi_0}=\tilde{H}^+\ket{\phi_-}$, which belongs to the tower constructed from $\ket{\phi_-}$.
In Sec. \ref{sec:spectral_analysis}, some momentum states appeared promising for exhibiting persistent revivals. We compute the time-evolution for all $\ket{\phi_k}$, as well as the suitable momentum boosted reference state from before (i.e. $\ket{-}^{\otimes 6}\ket{0+00}$ projected and normalized in the $k$-momentum sector). The results can be found in Fig. \ref{fig:all_momenta}. As predicted by the analysis in Sec. \ref{sec:spectral_analysis}, $k=0$ and $k=\pi/5$ exhibit the most prominent revivals and a broken Casimir that appears to remain well above the value obtained for the generic state.

\section{Conclusion\label{sec:conclusion}}

Quantum simulations of gauge theories hold the promise of unraveling novel non-equilibrium physics of gauge theories. Intriguingly, the system can remain in a non-thermal state for an arbitrary long time under unitary evolution. This can happen if the spectrum exhibits Quantum Many-Body Scars. Even in such cases, this is not enough. In order to observe the lack of thermalization, we need to choose the initial state properly, i.e., in such a way that it has a significant overlap with the Quantum Many-Body Scars of the Hamiltonian's spectrum. 

One mechanism that gives rise to exact revivals, i.e., the system returning periodically to its initial state, is the presence of a spectrum-generating algebra. The lack of thermalization can also arise when this spectrum-generating algebra holds only approximately. This is the case for the model studied here, a Quantum Link Model of spin-1 for a pure gauge theory on a ladder.
In order to identify potential initial states that can show approximate revivals, we compute the entanglement entropy and introduce the concept of a \emph{broken Casimir} invariant. By resolving the latter in different momentum sectors, we identified different initial states that lead to persistent revivals.

We leave questions regarding the fate of these states for long times and large volumes for future work. With the demonstration of the spectrum-generating algebra mechanism for the spin-1 ladder system, one may ask what happens once we consider the full 2d system, higher spins, or even three spatial dimensions. Those questions are more challenging due to the fast growth of the dimension of the Hilbert space. We hope to address these questions in small systems with exact diagonalization or, for larger systems, with additional analytical insight. 

\section{Acknowledgements}
We thank Debasish Banejee, Klemen Kersic, Kiryl Pakrouski and Zlatko Papic for enlightening discussions. This research was supported in part by grant NSF PHY-2309135 to the Kavli Institute for Theoretical Physics (KITP). MKM is grateful for the hospitality of Perimeter Institute where part of this work was carried out. Research at Perimeter Institute is supported in part by the Government of Canada through the Department of Innovation, Science and Economic Development and by the Province of Ontario through the Ministry of Colleges and Universities. This research was also supported in part by the Simons Foundation through the Simons Foundation Emmy Noether Fellows Program at Perimeter Institute.

\bibliographystyle{JHEP}
\bibliography{pos}

@article{banerjeeQuantumScarsZero2021,
  title = {Quantum {{Scars}} from {{Zero Modes}} in an {{Abelian Lattice Gauge Theory}} on {{Ladders}}},
  author = {Banerjee, Debasish and Sen, Arnab},
  year = 2021,
  month = jun,
  journal = {Physical Review Letters},
  volume = {126},
  number = {22},
  pages = {220601},
  issn = {0031-9007, 1079-7114},
  doi = {10.1103/PhysRevLett.126.220601},
  urldate = {2026-02-04},
  langid = {english}
}

@article{bernienProbingManybodyDynamics2017,
  title = {Probing Many-Body Dynamics on a 51-Atom Quantum Simulator},
  author = {Bernien, Hannes and Schwartz, Sylvain and Keesling, Alexander and Levine, Harry and Omran, Ahmed and Pichler, Hannes and Choi, Soonwon and Zibrov, Alexander S. and Endres, Manuel and Greiner, Markus and Vuleti{\'c}, Vladan and Lukin, Mikhail D.},
  year = 2017,
  month = nov,
  journal = {Nature},
  volume = {551},
  number = {7682},
  pages = {579--584},
  issn = {0028-0836, 1476-4687},
  doi = {10.1038/nature24622},
  urldate = {2026-02-04},
  langid = {english}
}

@article{biswasScarsProtectedZero2022,
  title = {Scars from Protected Zero Modes and beyond in \${{U}}(1)\$ Quantum Link and Quantum Dimer Models},
  author = {Biswas, Saptarshi and Banerjee, Debasish and Sen, Arnab},
  year = 2022,
  month = may,
  journal = {SciPost Physics},
  volume = {12},
  number = {5},
  pages = {148},
  issn = {2542-4653},
  doi = {10.21468/SciPostPhys.12.5.148},
  urldate = {2026-02-04}
}

@article{buddeQuantumManybodyScars2024,
  title = {Quantum Many-Body Scars for Arbitrary Integer Spin in 2+{{1D Abelian}} Gauge Theories},
  author = {Budde, Thea and Krsti{\'c} Marinkovi{\'c}, Marina and Barros, Joao C. Pinto},
  year = 2024,
  journal = {Phys. Rev. D},
  volume = {110},
  number = {9},
  pages = {094506},
  doi = {10.1103/PhysRevD.110.094506}
}

@article{bullQuantumScarsEmbeddings2020,
  title = {Quantum Scars as Embeddings of Weakly "Broken" {{Lie}} Algebra Representations},
  author = {Bull, Kieran and Desaules, Jean-Yves and Papic, Zlatko},
  year = 2020,
  month = apr,
  journal = {Physical Review B},
  volume = {101},
  number = {16},
  eprint = {2001.08232},
  primaryclass = {cond-mat, physics:quant-ph},
  pages = {165139},
  issn = {2469-9950, 2469-9969},
  doi = {10.1103/PhysRevB.101.165139},
  urldate = {2023-05-23},
  archiveprefix = {arXiv}
}

@article{chandranQuantumManyBodyScars2023,
  title = {Quantum {{Many-Body Scars}}: {{A Quasiparticle Perspective}}},
  shorttitle = {Quantum {{Many-Body Scars}}},
  author = {Chandran, Anushya and Iadecola, Thomas and Khemani, Vedika and Moessner, Roderich},
  year = 2023,
  month = mar,
  journal = {Annual Review of Condensed Matter Physics},
  volume = {14},
  number = {Volume 14, 2023},
  pages = {443--469},
  publisher = {Annual Reviews},
  issn = {1947-5454, 1947-5462},
  doi = {10.1146/annurev-conmatphys-031620-101617},
  urldate = {2025-10-29},
  langid = {english}
}

@article{cochranVisualizingDynamicsCharges2025,
  title = {Visualizing Dynamics of Charges and Strings in (2 + 1){{D}} Lattice Gauge Theories},
  author = {Cochran, T. A. and Jobst, B. and Rosenberg, E. and Lensky, Y. D. and Gyawali, G. and Eassa, N. and Will, M. and Szasz, A. and Abanin, D. and Acharya, R. and Aghababaie Beni, L. and Andersen, T. I. and Ansmann, M. and Arute, F. and Arya, K. and Asfaw, A. and Atalaya, J. and Babbush, R. and Ballard, B. and Bardin, J. C. and Bengtsson, A. and Bilmes, A. and Bourassa, A. and Bovaird, J. and Broughton, M. and Browne, D. A. and Buchea, B. and Buckley, B. B. and Burger, T. and Burkett, B. and Bushnell, N. and Cabrera, A. and Campero, J. and Chang, H.-S. and Chen, Z. and Chiaro, B. and Claes, J. and Cleland, A. Y. and Cogan, J. and Collins, R. and Conner, P. and Courtney, W. and Crook, A. L. and Curtin, B. and Das, S. and Demura, S. and De Lorenzo, L. and Di Paolo, A. and Donohoe, P. and Drozdov, I. and Dunsworth, A. and Eickbusch, A. and Elbag, A. Moshe and Elzouka, M. and Erickson, C. and Ferreira, V. S. and Burgos, L. Flores and Forati, E. and Fowler, A. G. and Foxen, B. and Ganjam, S. and Gasca, R. and Genois, {\'E}. and Giang, W. and Gilboa, D. and Gosula, R. and Grajales Dau, A. and Graumann, D. and Greene, A. and Gross, J. A. and Habegger, S. and Hansen, M. and Harrigan, M. P. and Harrington, S. D. and Heu, P. and Higgott, O. and Hilton, J. and Huang, H.-Y. and Huff, A. and Huggins, W. and Jeffrey, E. and Jiang, Z. and Jones, C. and Joshi, C. and Juhas, P. and Kafri, D. and Kang, H. and Karamlou, A. H. and Kechedzhi, K. and Khaire, T. and Khattar, T. and Khezri, M. and Kim, S. and Klimov, P. and Kobrin, B. and Korotkov, A. and Kostritsa, F. and Kreikebaum, J. and Kurilovich, V. and Landhuis, D. and {Lange-Dei}, T. and Langley, B. and Lau, K.-M. and Ledford, J. and Lee, K. and Lester, B. and Le Guevel, L. and Li, W. and Lill, A. T. and Livingston, W. and Locharla, A. and Lundahl, D. and Lunt, A. and Madhuk, S. and Maloney, A. and Mandr{\`a}, S. and Martin, L. and Martin, O. and Maxfield, C. and McClean, J. and McEwen, M. and Meeks, S. and Megrant, A. and Miao, K. and Molavi, R. and Molina, S. and Montazeri, S. and Movassagh, R. and Neill, C. and Newman, M. and Nguyen, A. and Nguyen, M. and Ni, C.-H. and Ottosson, K. and Pizzuto, A. and Potter, R. and Pritchard, O. and Quintana, C. and Ramachandran, G. and Reagor, M. and Rhodes, D. and Roberts, G. and Sankaragomathi, K. and Satzinger, K. and Schurkus, H. and Shearn, M. and Shorter, A. and Shutty, N. and Shvarts, V. and Sivak, V. and Small, S. and Smith, W. C. and Springer, S. and Sterling, G. and Suchard, J. and Sztein, A. and Thor, D. and Torunbalci, M. and Vaishnav, A. and Vargas, J. and Vdovichev, S. and Vidal, G. and Vollgraff Heidweiller, C. and Waltman, S. and Wang, S. X. and Ware, B. and White, T. and Wong, K. and Woo, B. W. K. and Xing, C. and Yao, Z. Jamie and Yeh, P. and Ying, B. and Yoo, J. and Yosri, N. and Young, G. and Zalcman, A. and Zhang, Y. and Zhu, N. and Zobrist, N. and Boixo, S. and Kelly, J. and Lucero, E. and Chen, Y. and Smelyanskiy, V. and Neven, H. and {Gammon-Smith}, A. and Pollmann, F. and Knap, M. and Roushan, P.},
  year = 2025,
  month = jun,
  journal = {Nature},
  volume = {642},
  number = {8067},
  pages = {315--320},
  issn = {0028-0836, 1476-4687},
  doi = {10.1038/s41586-025-08999-9},
  urldate = {2026-02-05},
  langid = {english}
}

@article{daiFourbodyRingexchangeInteractions2017,
  title = {Four-Body Ring-Exchange Interactions and Anyonic Statistics within a Minimal Toric-Code {{Hamiltonian}}},
  author = {Dai, Han-Ning and Yang, Bing and Reingruber, Andreas and Sun, Hui and Xu, Xiao-Fan and Chen, Yu-Ao and Yuan, Zhen-Sheng and Pan, Jian-Wei},
  year = 2017,
  month = dec,
  journal = {Nature Physics},
  volume = {13},
  number = {12},
  pages = {1195--1200},
  issn = {1745-2473, 1745-2481},
  doi = {10.1038/nphys4243},
  urldate = {2026-02-05},
  langid = {english}
}

@article{dalessioQuantumChaosEigenstate2016,
  title = {From Quantum Chaos and Eigenstate Thermalization to Statistical Mechanics and Thermodynamics},
  author = {D'Alessio, Luca and Kafri, Yariv and Polkovnikov, Anatoli and Rigol, Marcos},
  year = 2016,
  month = may,
  journal = {Advances in Physics},
  volume = {65},
  number = {3},
  pages = {239--362},
  issn = {0001-8732, 1460-6976},
  doi = {10.1080/00018732.2016.1198134},
  urldate = {2026-02-04},
  langid = {english}
}

@article{desaulesProminentQuantumManybody2023,
  title = {Prominent Quantum Many-Body Scars in a Truncated {{Schwinger}} Model},
  author = {Desaules, Jean-Yves and Hudomal, Ana and Banerjee, Debasish and Sen, Arnab and Papi{\'c}, Zlatko and Halimeh, Jad C.},
  year = 2023,
  journal = {Phys. Rev. B},
  volume = {107},
  number = {20},
  pages = {205112},
  doi = {10.1103/PhysRevB.107.205112}
}

@article{desaulesWeakErgodicityBreaking2023,
  title = {Weak Ergodicity Breaking in the {{Schwinger}} Model},
  author = {Desaules, Jean-Yves and Banerjee, Debasish and Hudomal, Ana and Papi{\'c}, Zlatko and Sen, Arnab and Halimeh, Jad C.},
  year = 2023,
  month = may,
  journal = {Physical Review B},
  volume = {107},
  number = {20},
  pages = {L201105},
  issn = {2469-9950, 2469-9969},
  doi = {10.1103/PhysRevB.107.L201105},
  urldate = {2026-02-04},
  langid = {english}
}

@article{fontanaQuantumSimulatorLink2023,
  title = {Quantum Simulator of Link Models Using Spinor Dipolar Ultracold Atoms},
  author = {Fontana, Pierpaolo and Barros, Joao C. Pinto and Trombettoni, Andrea},
  year = 2023,
  month = apr,
  journal = {Physical Review A},
  volume = {107},
  number = {4},
  pages = {043312},
  issn = {2469-9926, 2469-9934},
  doi = {10.1103/PhysRevA.107.043312},
  urldate = {2026-02-04},
  langid = {english}
}

@article{halimehRobustQuantumManybody2023,
  title = {Robust Quantum Many-Body Scars in Lattice Gauge Theories},
  author = {Halimeh, Jad C. and Barbiero, Luca and Hauke, Philipp and Grusdt, Fabian and Bohrdt, Annabelle},
  year = 2023,
  month = may,
  journal = {Quantum},
  volume = {7},
  pages = {1004},
  issn = {2521-327X},
  doi = {10.22331/q-2023-05-15-1004},
  urldate = {2026-02-05},
  langid = {english}
}

@article{hartseStabilizerScars2025,
  title = {Stabilizer {{Scars}}},
  author = {Hartse, Jeremy and Fidkowski, Lukasz and Mueller, Niklas},
  year = 2025,
  month = aug,
  journal = {Physical Review Letters},
  volume = {135},
  number = {6},
  pages = {060402},
  issn = {0031-9007, 1079-7114},
  doi = {10.1103/n5hb-l5p5},
  urldate = {2026-02-05},
  langid = {english}
}

@article{methSimulatingTwodimensionalLattice2025,
  title = {Simulating Two-Dimensional Lattice Gauge Theories on a Qudit Quantum Computer},
  author = {Meth, Michael and Zhang, Jinglei and Haase, Jan F. and Edmunds, Claire and Postler, Lukas and Jena, Andrew J. and Steiner, Alex and Dellantonio, Luca and Blatt, Rainer and Zoller, Peter and Monz, Thomas and Schindler, Philipp and Muschik, Christine and Ringbauer, Martin},
  year = 2025,
  month = apr,
  journal = {Nature Physics},
  volume = {21},
  number = {4},
  pages = {570--576},
  issn = {1745-2473, 1745-2481},
  doi = {10.1038/s41567-025-02797-w},
  urldate = {2026-02-05},
  langid = {english}
}

@article{moudgalyaExactExcitedStates2018,
  title = {Exact Excited States of Nonintegrable Models},
  author = {Moudgalya, Sanjay and Rachel, Stephan and Bernevig, B. Andrei and Regnault, Nicolas},
  year = 2018,
  month = dec,
  journal = {Physical Review B},
  volume = {98},
  number = {23},
  pages = {235155},
  issn = {2469-9950, 2469-9969},
  doi = {10.1103/PhysRevB.98.235155},
  urldate = {2026-02-26},
  langid = {english}
}

@article{moudgalyaQuantumManybodyScars2022,
  title = {Quantum Many-Body Scars and {{Hilbert}} Space Fragmentation: A Review of Exact Results},
  shorttitle = {Quantum Many-Body Scars and {{Hilbert}} Space Fragmentation},
  author = {Moudgalya, Sanjay and Bernevig, B Andrei and Regnault, Nicolas},
  year = 2022,
  month = jul,
  journal = {Reports on Progress in Physics},
  volume = {85},
  number = {8},
  pages = {086501},
  publisher = {IOP Publishing},
  issn = {0034-4885},
  doi = {10.1088/1361-6633/ac73a0},
  urldate = {2025-10-29},
  langid = {english}
}

@article{muellerQuantumComputingUniversal2025,
  title = {Quantum Computing Universal Thermalization Dynamics in a (2 + 1){{D Lattice Gauge Theory}}},
  author = {Mueller, Niklas and Wang, Tianyi and Katz, Or and Davoudi, Zohreh and Cetina, Marko},
  year = 2025,
  month = jul,
  journal = {Nature Communications},
  volume = {16},
  number = {1},
  pages = {5492},
  issn = {2041-1723},
  doi = {10.1038/s41467-025-60177-7},
  urldate = {2026-02-05},
  langid = {english}
}

@article{sauSublatticeScarsTwodimensional2024,
  title = {Sublattice Scars and beyond in Two-Dimensional {{U}} ( 1 ) Quantum Link Lattice Gauge Theories},
  author = {Sau, Indrajit and Stornati, Paolo and Banerjee, Debasish and Sen, Arnab},
  year = 2024,
  month = feb,
  journal = {Physical Review D},
  volume = {109},
  number = {3},
  pages = {034519},
  issn = {2470-0010, 2470-0029},
  doi = {10.1103/PhysRevD.109.034519},
  urldate = {2026-02-04},
  langid = {english}
}

@article{schecterManybodySpectralReflection2018,
  title = {Many-Body Spectral Reflection Symmetry and Protected Infinite-Temperature Degeneracy},
  author = {Schecter, Michael and Iadecola, Thomas},
  year = 2018,
  month = jul,
  journal = {Physical Review B},
  volume = {98},
  number = {3},
  pages = {035139},
  issn = {2469-9950, 2469-9969},
  doi = {10.1103/PhysRevB.98.035139},
  urldate = {2025-10-27},
  langid = {english}
}

@article{srednickiApproachThermalEquilibrium1999,
  title = {The Approach to Thermal Equilibrium in Quantized Chaotic Systems},
  author = {Srednicki, Mark},
  year = 1999,
  month = feb,
  journal = {Journal of Physics A: Mathematical and General},
  volume = {32},
  number = {7},
  pages = {1163--1175},
  issn = {0305-4470, 1361-6447},
  doi = {10.1088/0305-4470/32/7/007},
  urldate = {2026-02-04}
}

@article{yangPairingOffdiagonalLongrange1989,
  title = {{\emph{{$\eta$}}} Pairing and Off-Diagonal Long-Range Order in a {{Hubbard}} Model},
  author = {Yang, Chen Ning},
  year = 1989,
  month = nov,
  journal = {Physical Review Letters},
  volume = {63},
  number = {19},
  pages = {2144--2147},
  issn = {0031-9007},
  doi = {10.1103/PhysRevLett.63.2144},
  urldate = {2026-02-26},
  copyright = {http://link.aps.org/licenses/aps-default-license},
  langid = {english}
}

@article{yangSO4SYMMETRYHUBBARD1990,
  title = {{{SO}}{\textsubscript{4}} {{SYMMETRY IN A HUBBARD MODEL}}},
  author = {Yang, Chen Ning and Zhang, S.C.},
  year = 1990,
  month = jun,
  journal = {Modern Physics Letters B},
  volume = {04},
  number = {11},
  pages = {759--766},
  issn = {0217-9849, 1793-6640},
  doi = {10.1142/S0217984990000933},
  urldate = {2026-02-26},
  langid = {english}
}

@article{pakrouskiGroupTheoreticApproach2021a,
    title = {Group theoretic approach to many-body scar states in fermionic lattice models},
    volume = {3},
    doi = {10.1103/PhysRevResearch.3.043156},
    number = {4},
    journal = {Phys. Rev. Res.},
    author = {Pakrouski, Kiryl and Pallegar, Preethi N. and Popov, Fedor K. and Klebanov, Igor R.},
    year = {2021},
    note = {\_eprint: 2106.10300},
    pages = {043156},
}

@article{calajoQuantumManybodyScarring2025,
  title = {Quantum Many-Body Scarring in a Non-{{Abelian}} Lattice Gauge Theory},
  author = {Calaj{\'o}, Giuseppe and Cataldi, Giovanni and Rigobello, Marco and Wanisch, Darvin and Magnifico, Giuseppe and Silvi, Pietro and Montangero, Simone and Halimeh, Jad C.},
  year = 2025,
  month = mar,
  journal = {Physical Review Research},
  volume = {7},
  number = {1},
  pages = {013322},
  issn = {2643-1564},
  doi = {10.1103/PhysRevResearch.7.013322},
  urldate = {2026-03-06},
  langid = {english}
}

@article{geNonmesonicQuantumManyBody2024,
  title = {Nonmesonic {{Quantum Many-Body Scars}} in a {{1D Lattice Gauge Theory}}},
  author = {Ge, Zi-Yong and Zhang, Yu-Ran and Nori, Franco},
  year = 2024,
  month = jun,
  journal = {Physical Review Letters},
  volume = {132},
  number = {23},
  pages = {230403},
  issn = {0031-9007, 1079-7114},
  doi = {10.1103/PhysRevLett.132.230403},
  urldate = {2026-03-06},
  langid = {english}
}

@misc{guptaExactStabilizerScars2026,
  doi = {10.48550/ARXIV.2603.03062},
  url = {https://arxiv.org/abs/2603.03062},
  author = {Gupta,  Sabhyata and Sierant,  Piotr and Santos,  Luis and Stornati,  Paolo},
  title = {Exact stabilizer scars in two-dimensional $U(1)$ lattice gauge theory},
  publisher = {arXiv},
  year = {2026},
  copyright = {Creative Commons Attribution 4.0 International}
}

@article{miaoExactQuantumManyBody2025,
  title        = "Exact quantum many-body scars in {2D} quantum gauge models",
  author       = "Miao, Yuan and Li, Linhao and Katsura, Hosho and Yamazaki,
                  Masahito",
  year         =  2025,
  publisher = {arXiv},
  primaryClass = "cond-mat.str-el",
  eprint       = "2505.21921"
}

@inproceedings{Budde2026,
  series = {LATTICE2025},
  title = {Hilbert Space Fragmentation and Gauge Symmetry},
  url = {},
  DOI = {},
  author = {Budde, Thea and Marinkovic, Marina Krstic and Pinto Barros, Joao},
  year = {2026},
  pages = {125},
  collection = {LATTICE2025}
}

\end{document}